\documentclass {iopart}
\usepackage{graphicx}

\begin{document}

\title{Metal-insulator transition in Nd$_{1-x}$Eu$_{x}$NiO$_{3}$ compounds}

\author{M T Escote$^1$ , V B Barbeta$^2$,
R F Jardim$^3$ and J Campo$^4$}

\address{$^1$ Instituto de Qu\'{i}mica, Universidade Estadual de
S\~{a}o Paulo, Campus Araraquara, CP 353, 14801-970, Araraquara,
SP, Brazil}

\address{$^2$ Departamento de F\'{i}sica, Centro Universit\'{a}rio da
FEI, Av. Humberto de A. C. Branco 3972, 09850-901, S. B. Campo,
SP, Brazil}

\address{$^3$ Instituto de F\'{i}sica,
Universidade de S\~{a}o Paulo, CP 66318, 05315-970, S\~{a}o Paulo,
SP, Brazil}

\address{$^4$ Instituto de Ciencia de Materiales de Arag\'{o}n,
Facultad de Ciencias, Universidad de Zaragoza, C/ Pedro Cerbuna
12, 50009 Zaragoza, Spain}

\ead{rjardim@if.usp.br}

\begin{abstract}
Polycrystalline Nd$_{1-x}$Eu$_{x}$NiO$_3$ ($0 \leq x \leq 0.5$)
compounds were synthesized in order to investigate the character
of the metal-insulator (MI) phase transition in this series.
Samples were prepared through the sol-gel route and subjected to
heat treatments at $\sim$1000 $^\circ$C under oxygen pressures as
high as 80 bar. X-ray Diffraction (XRD) and Neutron Powder
Diffraction (NPD), electrical resistivity $\rho(T)$, and
Magnetization $M(T)$ measurements were performed on these
compounds. The results of NPD and XRD indicated that the samples
crystallize in an orthorhombic distorted perovskite structure,
space group $Pbnm$. The analysis of the structural parameters
revealed a sudden and small expansion of $\sim$$0.2$\% of the unit
cell volume when electronic localization occurs. This expansion
was attributed to a small increase of $\sim$0.003 \AA{} of the
average Ni-O distance and a simultaneous decrease of $\sim$$-
0.5^\circ$ of the Ni-O-Ni superexchange angle. The $\rho(T)$
measurements revealed a MI transition occurring at temperatures
ranging from $T_{\rm MI}\sim 193$ to 336 K for samples with $x =
0$ and 0.50, respectively. These measurements also show a large
thermal hysteresis in NdNiO$_{3}$ during heating and cooling
processes suggesting a first-order character of the phase
transition at $T_{\rm MI}$. The width of this thermal hysteresis
was found to decrease appreciably for the sample
Nd$_{0.7}$Eu$_{0.3}$NiO$_{3}$. The results indicate that cation
disorder associated with increasing substitution of Nd by Eu is
responsible for changing the first order character of the
transition in NdNiO$_{3}$.
\end{abstract}

\maketitle

\section{Introduction}

Since the discovery of high $T_{\rm c}$ superconductivity in
copper oxides a great interest has been renewed in
transition-metal oxide systems \cite{Bed}. The perovskites with
general formula \emph{R}NiO$_{3}$ (\emph{R} = rare earth) are
typical examples which have been investigated due to their
interesting transport and magnetic properties [2-15]. Moreover,
these compounds provide a remarkable opportunity to study the
relationship between structural changes and physical properties in
perovskite-like oxides \cite{Gar2,Med2,Zho1}. Perhaps the most
interesting property of these series is the occurrence of a
temperature-driven metal-insulator (MI) transition in compounds
with \emph{R}$\neq$La \cite{Lac, Tor, Xu}. In fact, electrical
resistivity measurements performed on polycrystalline and thin
films of LaNiO$_{3}$ revealed a metal-like behavior down to 1.5 K
\cite{Tor,Esc2}. However, in other members of this family, as
PrNiO$_{3}$ and NdNiO$_{3}$, a very sharp MI transition is
frequently verified \cite{Tor}. The temperature $T_{\rm MI}$, in
which the MI transition occurs, has been found to increase
systematically as the size of the rare earth ion decreases being
130 K for PrNiO$_{3}$ and reaching values as high as 480 K in
EuNiO$_{3}$ \cite{Tor}. These compounds crystallize, in the
itinerant-electron phase, in an orthorhombically distorted
perovskite structure (space group $Pbnm$). A higher distortion of
the perovskite structure is often observed when the rare-earth ion
size is decreased due to the tilt of the NiO$_{6}$ octahedra and a
consequent reduction of the superexchange Ni-O-Ni angle
\cite{Lac}. These observed features seem to indicate that the MI
transition is close related to both the degree of distortion of
the ideal perovskite structure and the Ni-O-Ni bond angle
\cite{Gar1, Gar2}.

High-resolution Neutron Powder Diffraction (NPD) experiments
performed on PrNiO$_{3}$, NdNiO$_{3}$, and SmNiO$_{3}$ compounds
have revealed that the MI transition is accompanied by a small
structural change on the unit cells \cite{Gar2}. The data also
indicate that the unit cell volume undergoes a subtle increase
when the system evolves toward the insulator regime due to a
slight increase in the average Ni-O distance \cite{Gar2}. This
effect induces additional tilts of the NiO$_{6}$ octahedra, which
also implies in a subtle decrease of the Ni-O-Ni bond angle.

Some recent results in smaller rare earth ions (\emph{R} = Ho, Y,
Er and Lu) suggested a change in the crystal symmetry, from
orthorhombic $Pbnm$ to monoclinic $P2_1/n$, when the system
evolves from the metallic to the insulating state, due to a charge
disproportionation of Ni$^{3+}$ cations \cite{Alo2}. Results of
electron diffraction and Raman scattering indicated a symmetry
break also for NdNiO$_{3}$ \cite{Zag}. A direct observation of
charge order in epitaxial NdNiO$_{3}$ films, using resonant X-ray
scattering, was also reported \cite{Sta}. These results have led
to the proposition that charge disproportionation and the
structural phase transition from orthorhombic $P{bnm}$ to
monoclinic $P2_1/n$ should occur for all the rare-earth family at
$T_{\rm MI}$, although it would be difficult to discriminate the
two types of Ni sites. This charge ordered state would also
explain the unusual propagation vector observed in neutron
experiments, and it would make models with orbital order obsolete
\cite{Sta}. On the other hand, M\"{o}ssbauer spectroscopy
measurements in the Nd$_{0.98}$Fe$_{0.02}$NiO$_{3}$ compound,
performed by Presniakov {\it et al} \cite{Pre}, have showed the
existence of only one Ni chemical specimen, indicating that there
is no charge disproportionation in the insulating phase and,
therefore, there is no change in the lattice symmetry for this
compound at $T_{\rm MI}$.

Concerning the magnetic properties of these nickelates, muon-spin
relaxation experiments carried out on \emph{R}NiO$_{3}$
perovskites (\emph{R} = Pr, Nd, Sm, Eu) \cite{Tor} have revealed
the occurrence of an antiferromagnetic ordering of the Ni$^{3+}$
sub-lattice. For Pr and Nd compounds, the magnetic ordering occurs
at temperatures $T_{\rm N}$ close to $T_{\rm MI}$, whereas other
members of this family display $T_{\rm N} < T_{\rm MI}$. Such a
behavior has been confirmed by neutron powder diffraction
experiments \cite{Gar3}. The magnetic ordering of the Ni
sub-lattice exhibits an unusual antiferromagnetic order
\cite{Gar4} that has been interpreted as the result of a spin
density wave (SDW) phase \cite{Med2}. Furthermore, a combination
of transport and thermal characterizations have suggested that the
character of the MI transition is of first order in compounds that
display $T_{\rm N} \sim T_{\rm MI}$, and of second order for
compounds with $T_{\rm N} < T_{\rm MI}$ \cite{Zho2}. In fact,
high-resolution PES photoemission measurements by Vobornik {\it et
al} \cite{Vob} have revealed two different electronic regimes for
the \emph{R}NiO$_{3}$ compounds with $T_{\rm N} \sim T_{\rm MI}$
and $T_{\rm N} < T_{\rm MI}$. However, a more profound discussion
regarding this point requires consideration about the mechanism
responsible for the MI transition, which is still a point of
controversy in these nickelates. In fact, a complete understanding
of the MI transition in this series certainly requires extra
experimental data.

Within this scenario, this work focuses on the preparation and
characterization of polycrystalline samples of
Nd$_{1-x}$Eu$_{x}$NiO$_{3}$ ($0 \leq x \leq 0.5$). The
interrelations between the structural and physical properties of
these compounds have been studied. The results have enabled the
discussion of the changes observed in compounds where the
antiferromagnetic ordering occurs essentially at the same
temperature $T_{\rm N} \sim 200$ K but the MI transition varies
systematically from $\sim$$193$ to 336 K. In addition, the change
of the character of the MI transition is discussed within the
framework of the critical phenomena scenario from first to second
order when the Eu content is increased.

\section{Experimental Procedure}
\subsection{Samples Preparation}
Polycrystalline samples of Nd$_{1-x}$Eu$_{x}$NiO$_{3}$ ($0 \leq x
\leq 0.5$) were prepared by using sol-gel precursors, sintered at
high temperatures ($\sim$1000 $^\circ$C), and under oxygen
pressures up to 80 bar. Details of the route employed and the
sintering process for preparing these samples are described
elsewhere \cite{Esc3}. Samples of NdAlO$_{3}$ and EuAlO$_{3}$ were
also prepared in order to subtract the contribution to the
magnetic susceptibility $\chi(T)$ arising from the
Nd$_{1-x}$Eu$_{x}$-ions. These samples were produced by mixing
appropriate amounts of Nd$_{2}$O$_{3}$, Eu$_{2}$O$_{3}$, and
Al$_{2}$O$_{3}$. The intimate mixtures were sintered at
$\sim$$1400$ $^\circ$C in air for $\sim$$100$ h.

\subsection{Samples characterization}
All samples were characterized by means of X-ray Diffraction (XRD)
measurements in a Brucker D8 Advanced diffractometer using Cu
$K_{\alpha}$ radiation ($\lambda = 1.540\,56$ \AA). Typical
$2\theta$ angular scans ranging between 20 and $100^\circ$, in
steps of $0.02^\circ$, and accumulation time $\sim$10 s, were used
in these experiments. Data were collected at room temperature (RT)
and MgO was used as an internal standard. The cell parameters were
calculated from the corrected peak positions for all identified
reflections between $20^\circ \leq 2\theta \leq 100^\circ$.

Structural characterizations were also made by means of Neutron
Powder Diffraction (NPD) measurements in two selected samples:
NdNiO$_{3}$ and Nd$_{0.7}$Eu$_{0.3}$NiO$_{3}$. The experiments
were performed in the high-flux and medium resolution D20
diffractometer at the Institut Laue-Langevin (Grenoble). The D20
instrument is equipped with a PSD-detector spanning an angular
range from 1 to $160^\circ$ ($2\theta$) with a wavelength $\lambda
= 1.2989$ \AA{} (Cu(200) monochromator). An "orange" cryostat was
used for the measurements performed at low temperatures. The data
were taken after first cooling the sample from RT down to
$\sim$$50$ K. Then the sample was warmed up (10 min/K) and the NPD
data were collected at several temperatures up to 320 K. The
accumulation time per spectra was 10 min and the temperature
stability is estimated as being close to 0.5 K. In order to
minimize the neutron absorption by Eu, the
Nd$_{0.7}$Eu$_{0.3}$NiO$_{3}$ sample was placed in a hollow
vanadium can.

All the diffraction patterns were analyzed by the Rietveld method
using the FullProf program \cite{Car1}. The data for $2\theta  >
100^\circ$ were excluded in the refinements due to the low
resolution of the D20 diffractometer in this angular range.

The temperature dependence of the electrical resistivity $\rho(T)$
was measured by the standard dc four-probe method in the
temperature range from 77 to 400 K. Four copper electrical leads
were attached with Ag epoxy to gold film contact pads on bar
shaped samples, and the sample temperature was measured using a Pt
thermometer. The temperature $T_{\rm MI}$, in which the
metal-insulator transition occurs, was defined as the temperature
of the maximum in the $(1/\rho)(\rmd \rho/\rmd T)$ against $T$
curves taken upon heating.

Magnetization $M(T)$ measurements were taken in a commercial MPMS
superconducting quantum interference device (SQUID) magnetometer
from Quantum Design. Zero-Field-Cooled (ZFC) and Field-Cooled (FC)
runs were performed, with temperature ranging from 5 to 400 K and
under dc magnetic fields as high as 10 kOe.

\section{RESULTS AND DISCUSSION}
\subsection{Crystal Structure}

Figure 1 displays some of the XRD patterns of
Nd$_{1-x}$Eu$_{x}$NiO$_{3}$ compounds obtained at RT. The data
showed no extra reflections belonging to impurity phases. In
addition, the systematic observation of the peak related to the
K$_{\beta}$ radiation for the most intense reflection occurring
close to $2\theta \sim 30^\circ$ strongly indicates that the
samples have a high degree of crystallinity.

\begin{figure} [htp]
\centering
\includegraphics [width=0.8\textwidth] {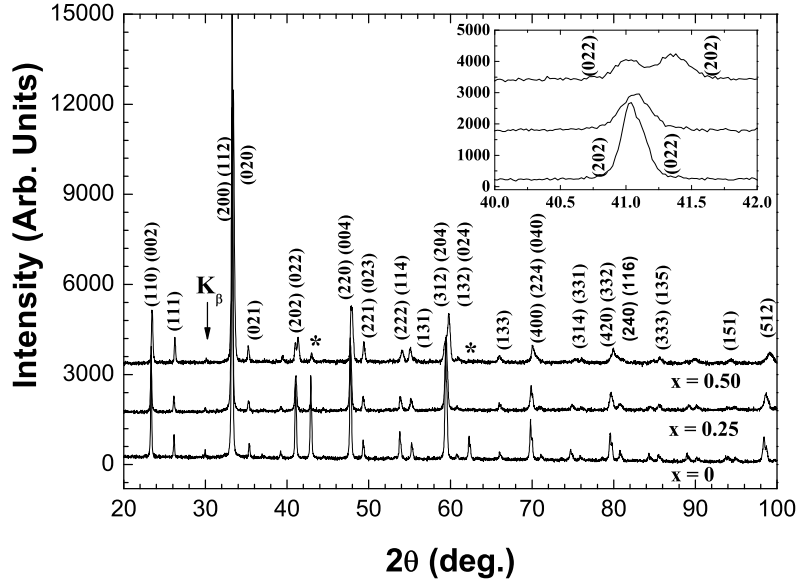}
\caption{\label{fig:epsart1} XRD patterns obtained at RT for
Nd$_{1-x}$Eu$_{x}$NiO$_{3}$, $x =$ 0, 0.25, and 0.50. The inset
displays the evolution of the peaks (022) and (202) for increasing
Eu content. The symbol "*" indicates the peaks corresponding to
the MgO internal standard.}
\end{figure}

All peaks of the X-ray patterns were indexed with the orthorhombic
distorted perovskite structure, space group $P{bnm}$ \cite{Dem}.
In this crystal structure, atoms are placed at the following
Wyckoff positions: $R$ ($R$ = Nd or Eu) and O(1) at 4c (x, y,
1/4); Ni at 4b (1/2, 0, 0); and O(2) at 8d (x, y, z). The actual
Nd substitution by Eu was inferred, for instance, by the
systematic changes observed in the (022) and (202) peaks with
increasing Eu content, as displayed in the inset of figure 1.  The
data first reveal a systematic shift of the reflections to higher
$2\theta$ values, as expected for a substitution of Nd by the
smaller Eu ion.  In addition, the separation of the peaks is
hardly seen for samples with $x = 0$ and 0.25, but there is a
clear broadening. For the sample with $x = 0.50$, both (022) and
(202) peaks are well separated in $2\theta$.

Structural refinements were performed in all diagrams by using as
starting parameters those reported in the literature for both
NdNiO$_{3}$ and EuNiO$_{3}$ compounds \cite{Gar2,Alo3}. The
refined cell parameters (\emph{a}, \emph{b}, and \emph{c}), the
unit cell volume (\emph{V}), and the atomic positions were found
to be in agreement with those expected for
Nd$_{1-x}$Eu$_{x}$NiO$_{3}$ samples ($0 \leq x \leq 0.5$), i. e.,
the structural parameters were between those of NdNiO$_{3}$ and
EuNiO$_{3}$ compounds. The resulting values of \emph{a}, \emph{b},
\emph{c}, and \emph{V} are plotted as a function of Eu
concentration in figure 2. As a general trend, \emph{V}, and the
cell parameters \emph{a} and \emph{c}, decrease as the Eu content
evolves, and an increase of the cell parameter \emph{b} is
observed. This is the expected behavior due to the smaller size of
the Eu ion compared to the Nd ion. For $x \cong 0.30$, a change in
the volume \emph{V} of the unit cell takes place and is certainly
related to the occurrence of the MI transition at RT.

\begin{figure} [htp]
\centering
\includegraphics [width=0.8\textwidth] {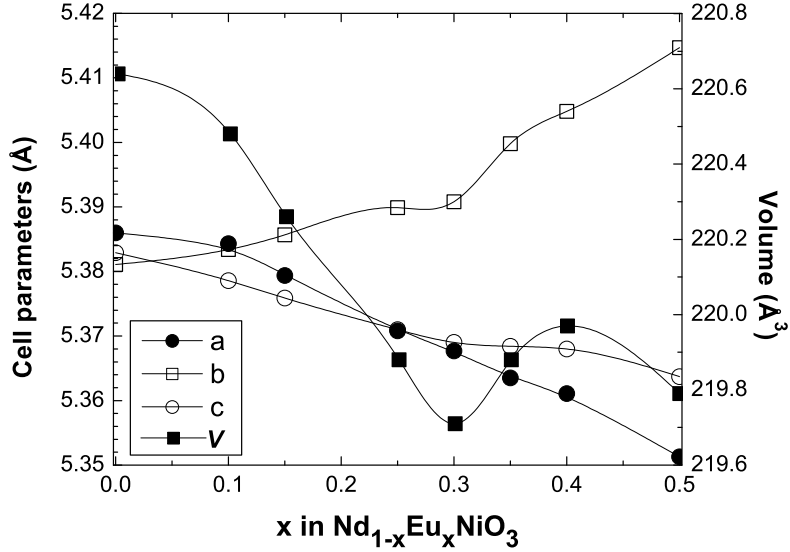}
\caption{\label{fig:epsart2} Room temperature cell parameters
\emph{a}, \emph{b}, and \emph{c}, and unit cell volume \emph{V} of
polycrystalline samples of Nd$_{1-x}$Eu$_{x}$NiO$_{3}$, as a
function of Eu content $x$, obtained through XRD data.}
\end{figure}

In order to study the thermal evolution of the crystal structure,
characterizations through NPD as a function of temperature were
carried out on two selected samples: NdNiO$_{3}$ and
Nd$_{0.7}$Eu$_{0.3}$NiO$_{3}$. An initial analysis of the NPD
patterns of both samples at RT revealed that the materials were
stoichiometric and single phase, as inferred from the absence of
peaks belonging to extra phases. These results agree with the
analysis performed previously through XRD.

All NPD diagrams were also analyzed through Rietveld refinements.
First, the RT patterns of Nd$_{1-x}$Eu$_{x}$NiO$_{3}$ samples were
analyzed and then the sequential FullProf program was used for all
other temperatures. In these refinements, the starting structural
parameters were those listed for NdNiO$_{3}$ \cite{Lac}.

A typical example of one of these refinements is shown in figure 3
that exhibits the NPD pattern for the sample of NdNiO$_{3}$ at $T
= 300$ K and the one calculated. This figure displays all the
reflections belonging to the desired phase which were indexed on
the basis of the orthorhombic GdFeO$_{3}$-type perovskite
\cite{Lac}. The excellent quality of the fitting was confirmed by
the $R_{\rm Bragg}$ reliability factor of $\sim$$3$, as displayed
in table 1. This table also contains the refined atomic positions,
cell parameters, factors $B$, the average Ni-O and $R$-O
distances, and the superexchange angle $\theta$ along with
reliability factors for both samples at three selected
temperatures.

The results of NPD at RT for the NdNiO$_{3}$ sample are in
excellent agreement with those previously reported for the same
compound \cite{Gar2}. For the Eu-substituted sample the results
are adequate, since they would be associated with a material
having cell parameters between those observed in NdNiO$_{3}$ and
EuNiO$_{3}$ compounds \cite{Gar2,Alo3}. The $R_{\rm Bragg}$ values
for the Nd$_{0.7}$Eu$_{0.3}$NiO$_{3}$ patterns, taken at several
temperatures, were greater than those for the unsubstituted
sample. Nevertheless, it is worth mentioning that excellent
Rietveld refinements in all NPD diagrams were obtained by adopting
the orthorhombic GdFeO$_{3}$-type perovskite. Obviously, due to
the increasing disorder related to the partial substitution of Nd
by Eu, the reliability factors $R_{\rm Bragg}$ for the
Eu-substituted sample were greater than those for the NdNiO$_{3}$
compound. However, the values of $R_{\rm Bragg}$ between 5 and 6
in Nd$_{0.7}$Eu$_{0.3}$NiO$_{3}$ are similar to the ones obtained
for the NdNiO$_{3}$ compound elsewhere \cite{Gar2}.

\fulltable{\label{tabl3}Structural parameters obtained via
Rietveld analysis for the NdNiO$_3$ and
Nd$_{0.7}$Eu$_{0.3}$NiO$_{3}$ compounds for three different
temperatures. The average interatomic distances and the
superexchange angle $\theta$ for the NiO$_6$ octahedra are also
displayed.} \br
Sample&\centre{3} {NdNiO$_3$} &\centre{3}{Nd$_{0.7}$Eu$_{0.3}$NiO$_{3}$ }\\
\ns \ns
&\crule{3}&\crule{3}\\
$T$(K)& \centre{1}{170} &\centre{1}{214} & \centre{1}{301} & \centre{1}{230} & \centre{1}{280} & \centre{1}{300}\\
\ns
\mr \emph{a}(\AA) & 5.395(1) & 5.394(2) & 5.399(2) & 5.365(3) & 5.362(3) & 5.363(2)\\
\emph{b}(\AA) & 5.383(1) & 5.377(1) & 5.380(1) & 5.401(2) & 5.398(2) & 5.396(2)\\
\emph{c}(\AA) & 7.616(2) & 7.612(2) & 7.617(2) & 7.615(3) & 7.612(3) & 7.612(3)\\
\emph{V}(\AA$^3$) & 221.16(1) & 220.79(2) & 221.25(2) & 220.64(3) & 220.32(3) & 220.25(3)\\
\mr $R$\\
x & 0.997(2) & 0.998(2) & 0.997(2) & 0.993(2) & 0.993(2) & 0.993(2)\\
y & 0.0370(6) & 0.0349(7) & 0.0345(7) & 0.0400(1) & 0.0397(1) & 0.0388(1)\\
$B$(\AA$^2$) & 0.56(6) & 0.14(6) & 0.21(6) & 0.30(8) & 0.32(9) & 0.34(9)\\
\mr Ni &  &  & &  &  & \\
$B$(\AA$^2$) & 0.15(4) & 0.18(4) & 0.22(4) & 0.22(6) & 0.26(6) & 0.25(6)\\
\mr O(1)\\
x & 0.076(2) & 0.074(2) & 0.075(2) & 0.073(2) & 0.072(2) & 0.073(2)\\
y & 0.490(1) & 0.491(1) & 0.491(1) & 0.483(2) & 0.484(2) & 0.484(2)\\
$B$(\AA$^2$) & 0.12(3) & 0.10(4) & 0.12(8) & 0.39(9) & 0.42(9) & 0.41(9)\\
\mr O(2)\\
x & 0.715(1) & 0.716(1) & 0.715(1) & 0.714(1) & 0.714(1) & 0.714(1)\\
y & 0.289(1) & 0.287(1) & 0.286(1) & 0.294(1) & 0.294(1) & 0.294(1)\\
z & 0.0353(9) & 0.0348(9) & 0.0342(9) & 0.0389(1) & 0.0390(1) & 0.0384(1)\\
$B$(\AA$^2$) & 0.41(6) & 0.41(6) & 0.53(8) & 0.35(9) & 0.38(9) & 0.40(9)\\
\mr $\langle{\rm Ni-O(1)}\rangle$(\AA) & 1.948(5) & 1.945(5) & 1.947(5) & 1.945(5) & 1.943(5) & 1.943(5)\\
\mr $\langle{\rm Ni-O(2)}\rangle$(\AA) & 1.945(5) & 1.940(5) & 1.942(5) & 1.950(5) & 1.949(5) & 1.948(5)\\
\mr $\langle{\rm Ni-O}\rangle$(\AA) & 1.946(5) & 1.942(5) & 1.944(5) & 1.949(5) & 1.947(5) & 1.946(5)\\
\mr $\langle{R{\rm-O}}\rangle$(\AA) & 2.521(5) & 2.525(5) & 2.528(5) & 2.502(2) & 2.503(2) & 2.505(2)\\
\mr $\theta$ $(^\circ)$ & 156.5(1) & 157.0(1) & 157.1(1) & 155.2(2) & 155.8(2) & 155.6(2)\\
\mr $R_{\rm Bragg}$ & 2.8 & 2.9 & 3.0 & 6.0 & 5.4 & 5.7\\
\br
\endfulltable

\begin{figure}[htp]
\centering
\includegraphics [width=0.8\textwidth] {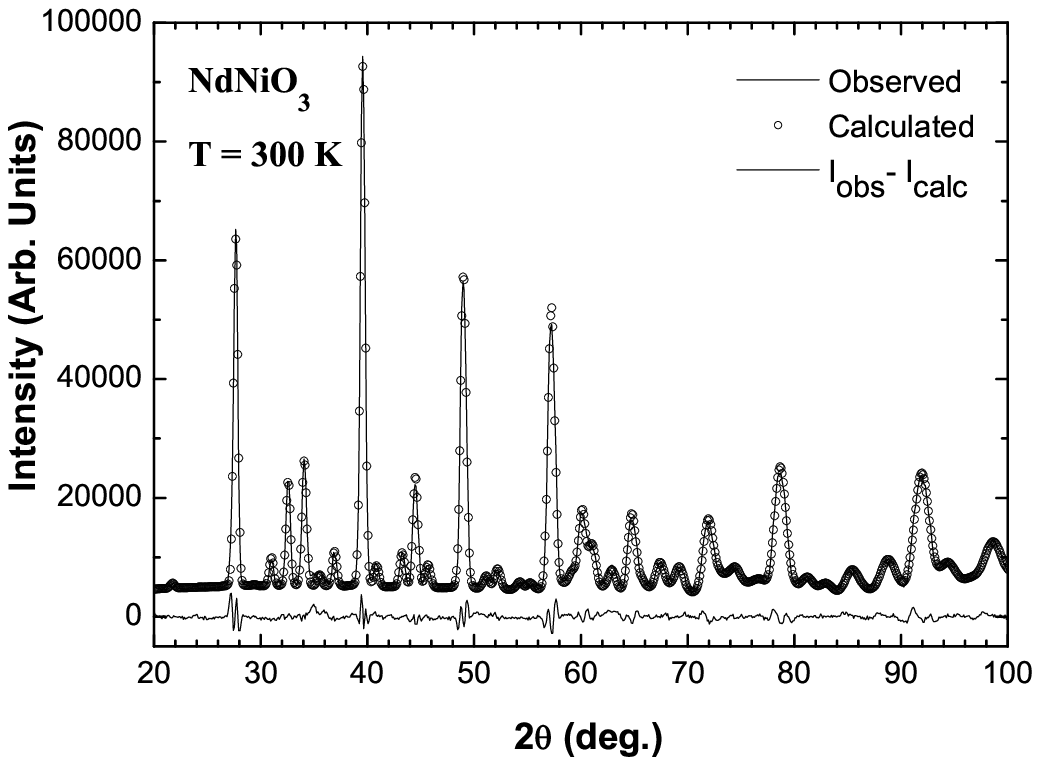}
\caption{\label{fig:epsart3} Observed (open circles), calculated
(solid line), and the difference profile (bottom line) of the room
temperature NPD pattern for the NdNiO$_{3}$ sample.}
\end{figure}

From figures 4 and 5, it is possible to observe a clear decrease
of the cell parameters \emph{a} and \emph{c}, and an increase of
the cell parameter \emph{b} with increasing Eu content. This
behavior, as expected due to the nature of the substitution, is
more pronounced in the unit cell volume which exhibits a decrease
from $\sim$221 \AA$^{3}$ just above $T_{\rm MI}$ for NdNiO$_{3}$
to $\sim$$220$ \AA$^{3}$ in Nd$_{0.7}$Eu$_{0.3}$NiO$_{3}$. It is
also observed that the Ni-O-Ni angle $\theta$ decreases and the
Ni-O distance increases with increasing Eu substitution (figure
6).

\begin{figure}[htp]
\centering
\includegraphics [width=0.8\textwidth] {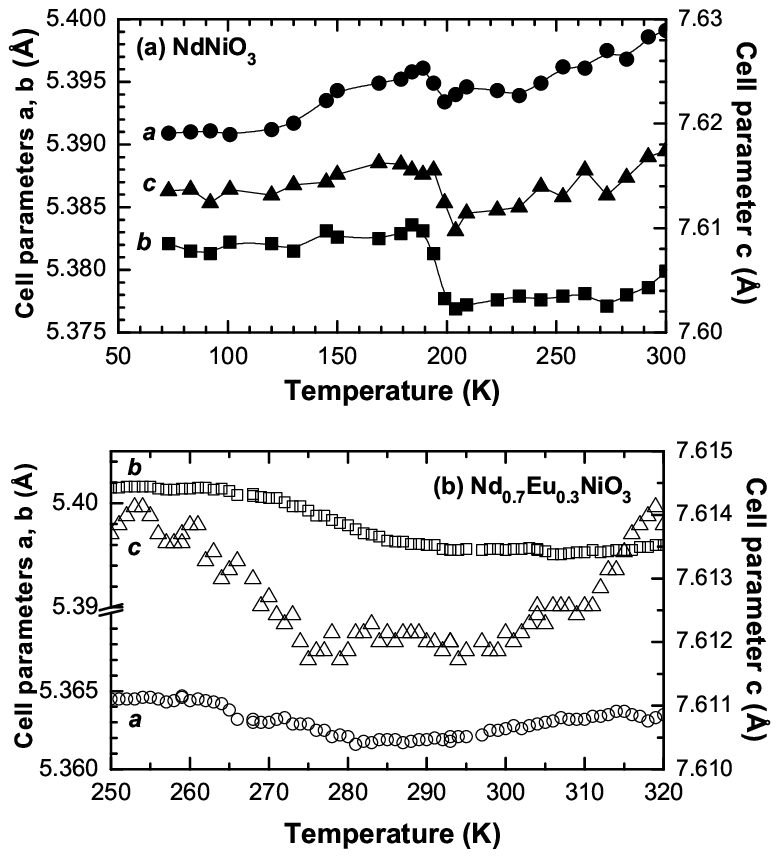}
\caption{\label{fig:epsart4} Temperature dependence of cell
parameters \emph{a}, \emph{b}, and \emph{c} for (a) NdNiO$_3$ and
(b) Nd$_{0.7}$Eu$_{0.3}$NiO$_{3}$ showing the thermal expansion
across the MI transition.  The data were taken from NPD
measurements upon warming.}
\end{figure}

Figures 4(a) and 4(b) show a small discontinuity of the cell
parameters at temperatures where the MI transition takes place at
$\sim$$200$ K and $\sim$$270$ K for NdNiO$_{3}$ and
Nd$_{0.7}$Eu$_{0.3}$NiO$_{3}$, respectively. The changes in the
cell parameters \emph{b} and \emph{c} at $T_{\rm MI}$ were
noticeable and accounted for $\sim$$0.1$\% and 0.06\%,
respectively. On the other hand, the observed change in the
magnitude of the cell parameter \emph{a} at $T_{\rm MI}$ was found
to be much smaller, yielding $\sim$$0.03$\%.

These results are similar to those found in PrNiO$_{3}$,
NdNiO$_{3}$, and SmNiO$_{3}$ by Garc\'{i}a-Mu\~{n}oz {\it et al}
\cite{Gar2}, except for the case of the lattice parameter \emph{a}
in NdNiO$_{3}$. They found a decrease in the lattice parameter
\emph{a} when the transition to the insulating phase takes place.
Our results for this sample seems to be more accurate, and agree
with those obtained by Lacorre {\it et al} \cite{Lac} in samples
of both PrNiO$_{3}$ and NdNiO$_{3}$.

Within our experimental resolution, no vestiges of change in the
crystal symmetry (orthorhombic, space group $P{bnm}$) across the
MI transition has been observed for both samples. This points out
for a kind of isomorphic transition, as frequently attributed to
\emph{R}NiO$_{3}$ (\emph{R} = Nd and Pr) compounds at $T_{\rm MI}$
\cite{Gar2}. The absence of a change in the crystal symmetry
across the MI transition in Nd-based nickelates is in agreement
with recent M\"{o}ssbauer spectroscopy data in
NdNi$_{0.98}$Fe$_{0.02}$O$_{3}$ \cite{Pre}. The authors could not
find any evidence of more than one nickel chemical specie,
therefore indicating the absence of charge disproportionation in
the light Fe-substituted NdNiO$_{3}$ compound. Such a result
strongly suggests crystallographic equivalent Ni sites below
$T_{\rm MI}$ in NdNiO$_{3}$.

Figure 5(a) exhibits a smooth thermal contraction of the unit cell
volume with decreasing temperature for the NdNiO$_{3}$ compound.
An abrupt and small increase of this value close to the MI
transition temperature $T_{\rm ND} \sim 193$ K ($\sim$$T_{\rm
MI}$) is also observed. Similar behavior occurs for the
Nd$_{0.7}$Eu$_{0.3}$NiO$_{3}$ sample, as seen in figure 5(b). In
this case, the change in the unit cell volume takes place at
$T_{\rm ND} \sim$ 273 K, a temperature similar to $T_{\rm MI} \sim
270$ K obtained from the $\rho(T)$ data. The unit cell volume
expansions were estimated to be $\Delta V/V_{0} \sim$ 0.22\% and
0.18\% for $x = 0$ and 0.30, respectively. The result for $x = 0$
is in excellent agreement with the value of $\Delta V/V_{0} = $
0.23\% reported previously \cite{Gar2}. The value of $\Delta
V/V_{0} = $ 0.18\%, for the sample with $x = 0.30$, is in line
with the expected decrease of $\Delta V/V_{0}$ that is observed
when the ionic radius is decreased.

The partial substitution of Nd by Eu also results in an
appreciable change of the temperature width $\Delta T$ in which
the MI transition occurs. This is clearly observed from the
temperature dependence of the unit cell volume across the MI
transition. From the data shown in figure 5, we have estimated
$\Delta T \sim 15$ and 35 K for samples with $x = 0$ and 0.30,
respectively. This remarkable difference in  $\Delta T$ points to
a change in the nature of the phase transition in this series.

\begin{figure}[htp]
\centering
\includegraphics [width=0.8\textwidth] {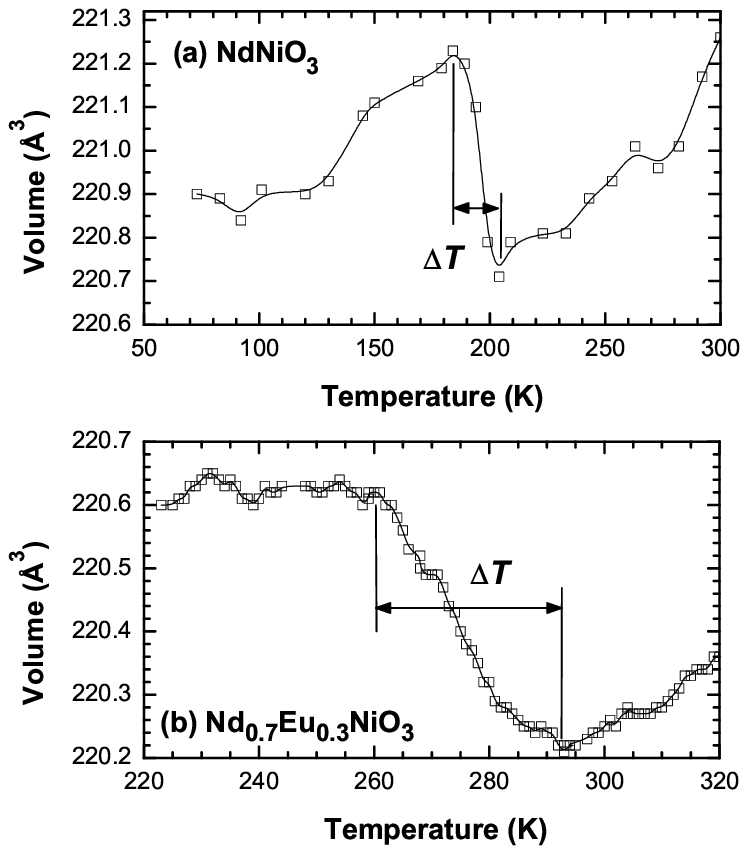}
\caption{\label{fig:epsart5} Temperature dependence of the unit
cell volume \emph{V} for (a) NdNiO$_3$ and
(b)Nd$_{0.7}$Eu$_{0.3}$NiO$_{3}$. The data were taken from NPD
measurements upon warming.}
\end{figure}

Changes in the cell parameters and a small increase in the unit
cell volume ($\Delta V/V_{0} \sim 0.2$\%) when the insulating
phase is established were detected. These changes are certainly
related to the increase in the Ni-O bond length induced by
electronic localization. The small change in the unit cell volume
is in agreement with the ten-times smaller change of $\Delta
V/V_{0} \sim 0.02$\% observed in BaVS$_{3}$, a compounds that
undergoes an isomorphic MI transition \cite{Graf}. On the other
hand, the widely studied V$_{2}$O$_{3}$ system exhibits a much
larger change in the unit cell volume at the MI transition
($\sim$$3.5$\%), a feature that is accompanied by an actual change
in the lattice symmetry from monoclinic to trigonal \cite{Goo2}.
In the latter case, the MI transition seems to be more
complicated, due to the role played by the structural changes in
driving the MI transition. In any event, it seems that in the
series studied here the relatively small changes in the unit cell
volume are in fact induced by the electronic localization across
the MI transition and not the driving mechanism for the
transition.

Table 2 displays the thermal expansion coefficients $\alpha_{\rm
L}$ and $\alpha_{\rm V}$ calculated for both samples above $T_{\rm
MI}$. The first one was estimated from the linear temperature
dependence of the cell parameters \emph{a}, \emph{b}, and \emph{c}
above $T_{\rm MI}$, and the $\alpha_{\rm V}$ coefficients from the
combined $\alpha_{\rm L}$ values. These coefficients were also
obtained for the insulating phase (not shown) in a similar
fashion.  The thermal expansion coefficient related to the cell
parameter \emph{b}, $\alpha^{b}_{\rm L}$, was found to exhibit the
smallest overall thermal expansion, but the largest discontinuity
at the phase transition. These results seem to be in good
agreement with those already obtained by Garc\'{i}a-Mu\~{n}oz {\it
et al} \cite{Gar2} in both PrNiO$_3$ and SmNiO$_3$. In addition,
increasing Eu content resulted in an increase of both
$\alpha^{a}_{\rm L}$ and $\alpha^{c}_{\rm L}$ values and a
decrease of $\alpha^{b}_{\rm L}$. It is well known that the
variation of structural parameters could be a response for the
subtle change in the structural arrangement \cite{Gar2}. In fact,
the expansion of the unit cell at $T_{\rm MI}$, when the system
evolves to the insulating state, is accompanied by at least two
major changes in the unit cell parameters: an increase of the
average Ni-O distance $d_{\rm Ni-O}$, and a decrease of the
Ni-O-Ni bond-angle $\theta$.

\begin{table}
\caption{\label{tab:table2}Linear ($\alpha_{\rm L}$) and
volumetric ($\alpha_{\rm V}$) thermal expansion coefficients
calculated for the metallic phase in both compounds: NdNiO$_3$ ($x
= 0$) and Nd$_{0.7}$Eu$_{0.3}$NiO$_{3}$ ($x = 0.30$).}
\begin{indented}
\lineup
\item[]\begin{tabular}{ccccc}
\br
 $x$ & $\alpha^{a}_{\rm L}$ & $\alpha^{b}_{\rm L}$ &
 $\alpha^{c}_{\rm L}$ & $\alpha_V$ \\
& ($10^{-6}$)K$^{-1}$ &($10^{-6}$)K$^{-1}$&
($10^{-6}$)K$^{-1}$&($10^{-5}$)
 K$^{-1}$\\
\hline 0& 13.2 & 4.3 & 8.8 & 2.62 \\
0.30& 14.7 & 1.8 & 12.9 & 2.48 \\

\br
\end{tabular}
\end{indented}
\end{table}

The average Ni-O distance ($d_{\rm Ni-O}$) and the superexchange
angle ($\theta$) were thus calculated in order to verify the
structural changes across the MI transition for the two samples
analyzed. The temperature dependence of $d_{\rm Ni-O}$ and
$\theta$ obtained for both samples are depicted in figure 6. The
temperature dependence of the $d_{\rm Ni-O}$ across the MI
transition exhibits similar behavior for both compounds. A small
expansion of $d_{\rm Ni-O}$ at $T_{\rm MI}$ was verified when the
temperature is decreased. The changes in the
$\langle$Ni-O$\rangle$ distance were estimated to be $\Delta
d_{\rm Ni-O} \sim 0.003$ \AA{} for both compounds, a value
comparable to those found in NdNiO$_{3}$ and PrNiO$_{3}$
\cite{Gar2}. However, it is worth mentioning that the change in
$d_{\rm Ni-O}$ at $T_{\rm MI}$ for the sample with $x = 0$ is much
sharper in temperature than for the sample with $x = 0.30$. The
$\langle$Ni-O$\rangle$ distance has been estimated to be $d_{\rm
Ni-O} \sim 1.94$ \AA, a value that is in excellent agreement with
previous results for NdNiO$_{3}$ \cite{Gar2}.

\begin{figure}[htp]
\centering
\includegraphics [width=0.8\textwidth] {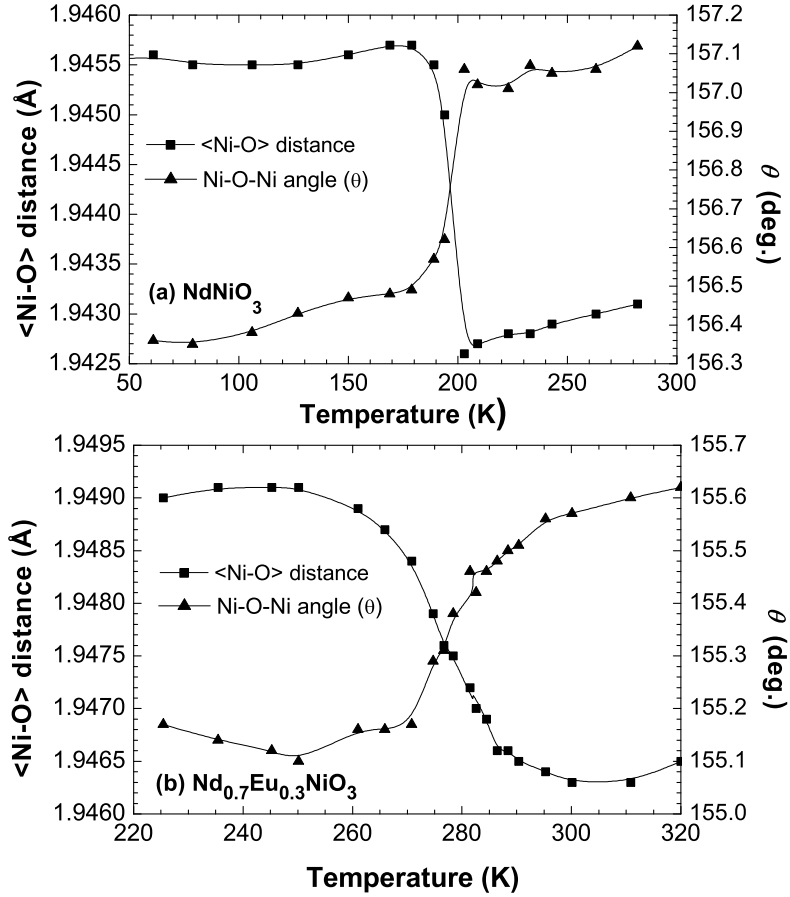}
\caption{\label{fig:epsart6} Average Ni-O distance
$\langle$Ni-O$\rangle$ and Ni-O-Ni superexchange angle ($\theta$)
for samples of (a) NdNiO$_{3}$ and (b)
Nd$_{0.7}$Eu$_{0.3}$NiO$_{3}$ obtained from Rietveld refinements
of the NPD data.}
\end{figure}

The temperature dependence of the superexchange angle $\theta$
also exhibits a smooth contraction $\Delta\theta$ close to $T_{\rm
MI}$ with decreasing temperature, as usually observed in other
nickelates \cite{Lac, Tor}. An estimate of this angular change at
$T_{\rm MI}$ yielded values of $\Delta\theta \sim - 0.5$ and $-
0.4^\circ$ for the samples of NdNiO$_{3}$ and
Nd$_{0.7}$Eu$_{0.3}$NiO$_{3}$, respectively. However, these
changes in $\Delta\theta$ across the MI transition can be
estimated by considering that such a change is due solely to
steric effects, or more appropriately, to the decrease of the
average Ni-O distance at $T_{\rm MI}$ \cite{Gar2}.

Within this context, it is possible to relate the degree of
distortion of the ideal perovskite structure, characterized by the
tolerance factor $t$, along with the superexchange angle $\theta$.
The tolerance factor is defined as

\begin{eqnarray}
t =\frac{d_{\rm R-O}}{d_{\rm Ni-O}\sqrt{2}}\label{eq1}
\end{eqnarray}

\noindent where $d_{\rm R-O}$ is the average (Nd/Eu)-O distance
and $d_{\rm Ni-O}$ is the average Ni-O distance. Therefore, a
linear approximation which correlates $\theta$ and $t$ resulted in
$\theta \sim 267.3t$. Differentiating this relation and the
expression for the tolerance factor, and combining both results,
the expected variation of the $\theta$ angle ($\Delta\theta$)
across the MI transition is given by

\begin{eqnarray}
\Delta\theta=-267.3\frac{d_{\rm R-O}}{d^2_{\rm
Ni-O}\sqrt{2}}\Delta d_{\rm Ni-O}
\end{eqnarray}

\noindent where $\Delta d_{\rm Ni-O}$ is the variation of $d_{\rm
Ni-O}$ at temperatures close to $T_{\rm MI}$.

By using such an approximation, $\Delta\theta$ values of
$\sim$$-0.38$ and $-0.32^\circ$ were estimated for the samples
with $x = 0$ and 0.30, respectively. The very good agreement
between the measured and estimated values of $\Delta\theta$
indicates that the expansion in the structural and geometrical
parameters could be understood as a steric response to the
increase of the Ni-O distance caused by the electronic
localization occurring at $T_{\rm MI}$. The values here obtained
for both samples were slightly smaller than $\Delta\theta \approx
- 0.46^\circ$ obtained before for NdNiO$_{3}$ \cite{Gar2}, but
with the same sign and the same order of magnitude.

\subsection{Electrical Resistivity}

The most striking feature displayed by this series is the
occurrence of a metal-insulator (MI) transition in a wide range of
temperature. Such a transition is easily seen in the temperature
dependence of the electrical resistivity, $\rho(T)$, for the
samples of Nd$_{1-x}$Eu$_{x}$NiO$_{3}$ ($0 \leq x \leq 0.5$), as
shown in figure 7. The overall behavior of the $\rho(T)$ data is
essentially the same for all samples and they exhibit four
important features: (1) a continuous increase in the MI transition
temperature with increasing Eu content; (2) a metallic-like
behavior above $T_{\rm MI}$, with a linear temperature dependence
of $\rho(T)$; (3) a fairly rapid increase of $\rho(T)$ at $T_{\rm
MI}$, related to the temperature-driven MI transition; and (4) a
clear thermal hysteresis occurring in a temperature interval
$\Delta T_{\rm MI}$ close to $T_{\rm MI}$, which decreases
appreciably as $T_{\rm MI}$ increases.

\begin{figure}[htp]
\centering
\includegraphics [width=0.8\textwidth] {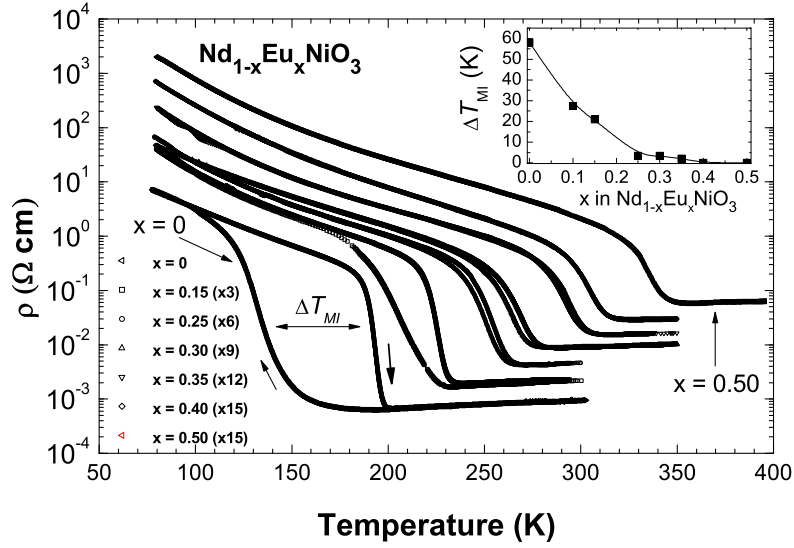}
\caption{\label{fig:epsart7} Temperature dependence of electrical
resistivity $\rho(T)$ of Nd$_{1-x}$Eu$_{x}$NiO$_{3}$ ($0 \leq x
\leq 0.5$) compounds. The data were taken during the heating and
the cooling processes, as indicated by the arrows.  The inset
displays the $\Delta T_{\rm MI}$ against $x$ curve.}
\end{figure}

The temperature in which the MI transition occurs is rather
sensitive to Eu concentration, as can be inferred from the data of
figure 7. The $\rho(T)$ data reveal that $T_{\rm MI}$ increases
continuously with increasing Eu concentration, ranging from
$\sim$$193$ K in NdNiO$_{3}$ to 336 K in
Nd$_{0.5}$Eu$_{0.5}$NiO$_{3}$ sample. The computed $T_{\rm MI}$
values are displayed in table 3. The NdNiO$_{3}$ sample was found
to exhibit $T_{\rm MI} \sim 193$ K, a value in excellent agreement
with others listed in literature \cite{Med1}. Increasing Eu
content resulted in a systematic increase of $T_{\rm MI}$, a
behavior consistent with intermediary values of $T_{\rm MI}$
ranging between $T_{\rm MI} \sim 196$ K for NdNiO$_{3}$ and
$T_{\rm MI} \sim 480$ K for EuNiO$_{3}$ \cite{Med1}.

The $\rho(T)$ data in the metallic regime, $T > T_{\rm MI}$, was
described as $\rho(T) = \rho_{0} + AT$ for all samples studied.
Such a behavior of $\rho(T)$ is typical of electron-phonon
scattering where $\rho_{0}$ is the residual electrical resistivity
and $A$ is related to the electron-phonon coupling constant
$\lambda_{\rm tr}$. Assuming that the linear dependence of the
electrical resistivity is due entirely to electron-phonon
scattering, then $\lambda_{\rm tr}$ can be obtained by using
\cite{Esc4}

\begin{eqnarray}
\lambda_{\rm tr}=0.246(\hbar\omega_{\rm p})^{2}A.
\end{eqnarray}

Considering the plasmon energy $\hbar\omega_{\rm p}\approx 1$ eV,
the same value obtained for LaNiO$_{3}$ \cite{Kem}, then
$\lambda_{\rm tr}$ can be estimated for all samples. It is also
possible to estimate the mean free path $l$ by doing

\begin{eqnarray}
l = \frac{4.95{\rm x}10^{-4}v_{\rm F}}{(\hbar\omega_{\rm
p})^{2}\rho}
\end{eqnarray}

\noindent where  $v_{\rm F}$ (Fermi velocity) was assumed to be of
the same magnitude of similar oxides ($v_{\rm F}\sim
2.2$x$10^{-4}$ cm/s for La$_{1.825}$Sr$_{0.175}$CuO$_4$)
\cite{Gur}.

From fits of $\rho(T)$ data we have obtained rising values of both
$\rho_{0}$ and $A$ with increasing Eu content. All these values,
as well as the values of $\lambda_{\rm tr}$, and $l$ for all
samples, are summarized in table 3. The mean free path $l$ was
estimated for a temperature just above the transition temperature
$T_{\rm MI}$. The values of  $\rho_{0}$, and $A$ are of the same
magnitude of those found in other related nickelates as, for
instance, $\rho_{0} = 4.6$x$10^{-4}$ $\Omega$cm and $A =
2.7$x$10^{-6}$ $\Omega$cm/K obtained in SmNiO$_{3}$ \cite{Cac}. It
is important to mention that the values of $\rho_{0}$ are expected
to be dependent on the heat treatments during the sample
preparation as well as on the oxygen nonstoichiometry
\cite{Esc3,Nik}.

\begin{table}
\caption{\label{tab:table3}Values of $T_{\rm MI}$, $\rho_0$, $A$,
$\lambda_{\rm tr}$ and $l$ for Nd$_{1-x}$Eu$_{x}$NiO$_{3}$.
$T_{\rm MI}$ was obtained from $(1/\rho)(\rmd \rho/\rmd T)$
against $T$ curves, and $\rho_0$ and $A$ were obtained through the
linear fitting of the experimental data in the metallic regime.
The mean free path $l$ values were estimated for a temperature
just above the MI transition.}
\begin{indented}
\lineup
\item[]\begin{tabular}{cccccc}
\br

$x$ & $T_{\rm MI}$& $\rho_0$& $A$& $\lambda_{\rm tr}$ & $l$\\
&(K)&($10^{-4}$ $\Omega$cm) &($10^{-6}$ $\Omega$cm/K)& & (\AA)\\
\hline 0 & 193 & 0.82 & 1.70 & 0.42 & 16.0\\
0.10 & 215 & 1.04 & 1.30 & 0.32 & 19.7 \\
0.15 & 226 & 0.97 & 2.32 & 0.57 & 18.8\\
0.25 & 254 & 1.47 & 1.37 & 0.34 & 12.9\\
0.30 & 270 & 3.28 & 2.45 & 0.60 & 10.8\\
0.35 & 293 & 4.76 & 2.59 & 0.64 & 8.4\\
0.40 & 304 & 7.88 & 3.54 & 0.87 & 5.5\\
0.50 & 336 & 20.2 & 5.42 & 1.33 & 2.8\\
\br
\end{tabular}
\end{indented}
\end{table}

Just below $T_{\rm MI}$, the electrical resistivity jumps by three
or four orders of magnitude and the electrical resistivity of the
system is better described by an Arrhenius-type activation process
$\rho(T) = \rho_{\rm s}{\rm exp}(E_{\rm g}/k_{\rm B}T)$, where
$\rho_{\rm s}$ is the temperature independent electrical
resistivity, $E_{\rm g}$ is the energy gap, and $k_{\rm B}$ is the
Boltzmann constant. We have fitted the low temperature data and
obtained the activation energy $E_{\rm g} \sim 44$ and 73 meV for
NdNiO$_{3}$ and Nd$_{0.7}$Eu$_{0.3}$NiO$_{3}$ compounds,
respectively. The activation energy is similar to that obtained by
Granados {\it et al} \cite{Gra} (25-28 meV) in the same compound
with $x = 0$ , although their results showed a smooth curvature in
the lg$R$ against $1/T$ curves, indicating that a simple activated
behavior is only a rough approximation of the insulating phase.
There are indications that the activation energy is also strongly
dependent on the preparation process, since a small oxygen
nonstoichiometry affects deeply the behavior of the insulating
phase \cite{Nik}.

Examining further the $\rho(T)$ curves in figure 7, one can note
that the thermal hysteresis $\Delta T_{\rm MI}$ is noticeable only
in samples with $x < 0.30$. In fact, it occurs in a temperature
interval $\Delta T_{\rm MI}$ as large as 58 K in NdNiO$_{3}$ and
is much less pronounced, or nearly absent, for the compound
Nd$_{0.65}$Eu$_{0.35}$NiO$_{3}$ where $\Delta T_{\rm MI} \sim$ 2 K
(see inset of figure 7). The hysteretic behavior observed here, at
least for the parent compound, is certainly related to the
first-order character of the MI transition and reflects the
temperature interval $\Delta T_{\rm MI}$ in which both phases
metallic (disordered) and insulating (ordered) coexist
\cite{Med2,Gar3}. The $\Delta T_{\rm MI}$ against $x$ plot shown
in the inset of figure 7 also indicates that the width $\Delta
T_{\rm MI}$ of this thermal hysteresis approaches zero for Eu
content $x \sim 0.25$. This strongly suggest that the partial
substitution of Nd by Eu modifies the character of this transition
from first to second order in samples with Eu content higher than
$\sim$0.25. Similar behavior has been observed in
Nd$_{0.5}$Sm$_{0.5}$NiO$_{3}$ which exhibited absence of thermal
hysteresis in $\rho(T)$ measurements, and the MI transition was
characterized as a second order phase transition \cite{Zho2}. This
proposition has to be further explored since a decrease in $\Delta
T_{\rm MI}$ is also expected as $T_{\rm MI}$ increases.

It should be mentioned that Nikulin {\it et al} \cite{Nik} could
not observe any noticeable thermal hysteresis in electrical
resistivity measurements performed in SmNiO$_{3}$. This behavior
was explained as a result of the kinetics of the M-I phase
transition, since the phase transition for SmNiO$_{3}$ occurs at
temperatures much higher than that for NdNiO$_{3}$. The kinetics
of the phase transformation is believed to be much slower for the
compound with Nd than that for the compound with Sm, resulting in
a noticeable thermal hysteresis only for the NdNiO$_{3}$ compound.

Figure 8 shows the temperature dependence of the logarithmic
derivative of the electrical resistivity curves, $(1/\rho)(\rmd
\rho/ \rmd T)$ against $T$ taken upon heating. The peaks provide a
better means of defining the value of $T_{\rm MI}$, as opposed to
the simple inspection of the $\rho(T)$ curves \cite{Fis}. The MI
transition is sharper for the parent compound, a feature that
manifests itself in the width of the peak in $(1/\rho)(\rmd
\rho/\rmd T)$.

\begin{figure}[htp]
\centering
\includegraphics [width=0.8\textwidth] {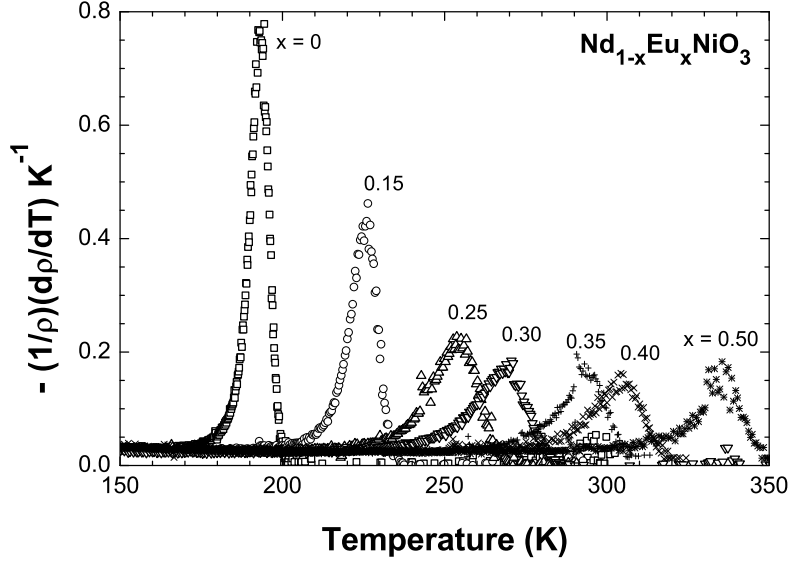}
\caption{\label{fig:epsart8} Temperature dependence of the
$(1/\rho)(\rmd \rho/\rmd T)$, obtained during heating process.}
\end{figure}

The broadening and the decrease of the peak intensity in
$(1/\rho)(\rmd \rho/\rmd T)$ with increasing Eu content suggest a
continuous transition from the metallic to the insulating state.
In addition, the trend shown in figure 8 indicates that increasing
substitution of Nd by Eu in NdNiO$_{3}$ moves the transition away
from a well defined value of $T_{\rm MI}$. This behavior is
certainly related to the increase of disorder which is also
responsible for the observed increase in the value of $\rho_0$
with increasing Eu content (see table 3). Such a partial
substitution is analogous to an increase of the quenched bond
randomness observed in systems as LaMnO$_3$ \cite{Dag}. In other
words, the coexistence between the ordered and disordered phases
at the phase transition disappears at a finite threshold amount of
Eu. In fact, neutron diffraction data have also suggested an
increase of the crystalline disorder in the
Nd$_{0.7}$Eu$_{0.3}$NiO$_{3}$ sample, as inferred from the
temperature interval in which the MI transition occurs (see
figures 5 and 6).

\subsection{Magnetization}

Magnetization measurements $M(T)$ were performed in all samples of
Nd$_{1-x}$Eu$_{x}$NiO$_{3}$ under magnetic fields $H$ up to 10
kOe. The magnetic susceptibility $\chi(T)$ was obtained by doing
$\chi(T) = M(T)M_m/Hm$, where $M_m$ is the molecular mass and $m$
is the mass of the sample.

In order to understand better the magnetic behavior of these
systems, specifically the behavior of the Ni sublattice which
orders antiferromagnetically, the magnetic contribution of the
rare earth ions to the susceptibility $\chi(T)$ was subtracted
from the original data. This subtraction was performed through
measurements of magnetization in samples with similar crystal
structure but without evidence for magnetic ordering of the metal.
In this case, samples of NdAlO$_{3}$ and EuAlO$_{3}$ were
prepared, where the Ni ion has been fully replaced by Al.

Once the corresponding $\chi(T)$ curves for the Al substituted
samples were obtained, they were fitted to a Curie-Weiss law in
the high temperature region ($T > 100$ K). Hence, it was possible
to obtain the magnetic susceptibility for each ion separately
(Nd$^{3+}$ and Eu$^{3+}$). Finally, the contribution of the Ni
ions (Ni$^{3+}$) to the total observed susceptibility $\chi(T)$
could be written as

\begin{eqnarray}
 \chi^{\rm Ni^{3+}} = \chi(T) -  \chi^{\rm Nd^{3+}}.(1-x) -
 \chi^{\rm Eu^{3+}}.(x).
\end{eqnarray}

Figure 9 displays a typical set of $\chi(T)$ curves after
performing the subtraction discussed above. The data belonging to
the samples with $x = 0$, 0.10, and 0.15 display a clear peak near
$\sim$$200$ K, corresponding to the antiferromagnetic ordering of
the Ni$^{3+}$ sublattice. For samples with $x = 0$ and $x = 0.10$,
$T_{\rm N} \sim 195$ K and for $x = 0.15$, $T_{\rm N} \sim 220$ K.

The curves of $\chi(T)$ for the samples where $T_{\rm N}$ is well
defined ($x = 0$, $x = 0.10$ and $x = 0.15$) were fit by the form
$\chi(T)= \chi_0 + C/(T-\theta_c)$, where $C$ is the Curie
constant, $\theta_c$ is the Curie-Weiss temperature and $\chi_0$
is a temperature-independent contribution to the susceptibility.
By using the Curie constant $C$ obtained in these fits, the values
obtained for $\mu_{\rm eff}$ were 1.75, 1.73, and 1.79 $\mu_{\rm
B}$ for the samples with $x = 0$, 0.10, and 0.15 respectively.
These values are in excellent agreement with $\mu_{\rm eff} \sim
1.76 \mu_{\rm B}$ expected for the free Ni$^{3+}$ ion at $T
> T_{\rm N}$.

\begin{figure}[htp]
\centering
\includegraphics [width=0.8\textwidth] {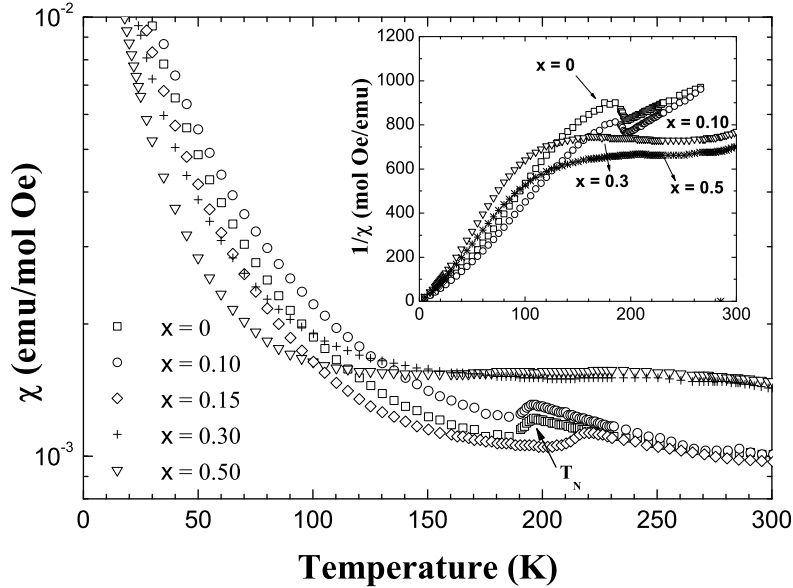}
\caption{\label{fig:epsart10} Curves of $\chi(T)$ for
Nd$_{1-x}$Eu$_{x}$NiO$_{3}$ ($0 \leq x \leq 0.5$) after the
subtraction of the contribution of the rare earth ions. The inset
shows the curves of $1/\chi(T)$.}
\end{figure}

The expected peak of the magnetic order of the Ni sublattice is
hardly seen in curves belonging to samples with $x \geq 0.3$. This
might be explained by the fact that increasing Nd substitution by
Eu results in a stronger effect of the Crystalline-Electric Field
(CEF). Therefore, higher values of $\chi(T)$ are obtained, hiding
the observation of the magnetic order of the Ni sublattice.
Besides, the magnetic contribution of the CEF of the Eu$^{3+}$ is
bigger than that of Nd$^{3+}$ ions, which makes the separation of
the Ni sublattice contribution more difficult than the one
performed above.

The magnitude of $\chi$ increases monotonically when the
temperature is decreased below $T_{\rm N}$. This indicates that
the thermal behavior of the Ni sublattice can not be understood
within the framework of a conventional antiferromagnet. One
possible explanation for this anomalous behavior would be the
presence of a canted-spin ferromagnetism. In fact, a small
irreversibility observed in our measurements in ZFC and FC curves
(not shown) would be an indication of weak ferromagnetism.
However, the irreversibility was observed to be almost field
independent, which seems to be inconsistent with the canted-spin
picture.

On the other hand, the low temperature behavior of the magnetic
susceptibility could be a result of two magnetic phases that
coexist below $T_{\rm N}$: one antiferromagnetic phase, where the
magnetic susceptibility decreases when the temperature decreases,
and a paramagnetic phase, where the magnetic susceptibility
increases further when the temperature decreases. This anomalous
behavior in the magnetic susceptibility has been reported
previously by Zhou {\it et al} \cite{Zho3} in measurements
performed in NdNiO$_{3}$ and Nd$_{0.5}$Sm$_{0.5}$NiO$_{3}$ and was
attributed to a charge disproportionation in the Ni sublattice,
resulting into alternating diamagnetic and paramagnetic Ni sites.
This point needs to be better clarified, and experiments are under
way to address that.

\section*{Conclusions}

In summary, we have produced high quality polycrystalline samples
of Nd$_{1-x}$Eu$_{x}$NiO$_{3}$ ($0 \leq x \leq 0.5$). From XRD and
NPD results we have found single-phase samples that crystallize in
the GdFeO$_{3}$-type orthorhombically distorted perovskite
structure (space group $Pbnm$). The thermal evolution of the
structural parameters across the MI transition revealed small unit
cell changes in temperatures close to the MI transition. Within
our experimental resolution, it was not possible to observe any
kind of structural phase transition at $T_{\rm MI}$ through the
analysis of the NPD data. The MI transition temperatures from NPD
($T_{\rm ND}$) were in good agreement with those found through
electrical resistivity ($T_{\rm MI}$) measurements.

The magnetic susceptibility data indicate, after subtracting the
magnetic contribution of the rare earth ions, that $\chi$
increases below $T_{\rm N}$ for all samples, suggesting a possible
coexistence of two phases: a paramagnetic phase and an
antiferromagnetic one. However, this point still needs to be
better clarified.

The substitution of Nd by Eu in the NdNiO$_3$ compound causes a
broadening in the variation of the unit cell volume across the MI
transition, which is also reflected in both the temperature
dependence of the Ni-O distance and the superexchange angle. It is
also observed an increase of $\rho_{0}$ in the electrical
resistivity curves, as the Eu content increases, and that the
transition becomes less defined. When the Eu content is higher
than $\sim$0.25, the thermal hysteresis observed in the electrical
resistivity curves, between the heating and cooling process, is
suppressed. These results indicate that crystalline disorder,
induced by the progressive substitution of Nd by Eu, is changing
the first-order character of the phase transition observed across
the MI boundary.

\ack The authors have benefited from the technical assistance of
Walter Soares de Lima. This work was supported by the Brazilian
agency Funda\c{c}\~{a}o de Amparo \`{a} Pesquisa do Estado de
S\~{a}o Paulo (FAPESP) under Grant No. 05/53241-9. One of us
(M.T.E.) acknowledges FAPESP fellow under Grant No. 97/11369-0 and
R.F.J. the Conselho Nacional de Desenvolvimento Cient\'{i}fico e
Tecnol\'{o}gico (CNPq) fellow under Grant No. 303272/2004-0.
Institut Laue Langevin (Grenoble) is acknowledged for the
allocated beam time on the D20 diffractometer.

\end{document}